 \documentclass[traditabstract]{aa} 
\usepackage{graphicx}
\usepackage{txfonts}
\usepackage{natbib}
\usepackage{color}

\begin{document}

   \title{Unveiling the physics behind the spectral variations of ``changing-look'' quasars with optical polarimetry}

   \titlerunning{Optical polarization of changing-look quasars}

   \author{F.~Marin\thanks{\email{frederic.marin@astro.unistra.fr}}}
   
   \institute{Universit\'e de Strasbourg, CNRS, Observatoire astronomique de Strasbourg, UMR 7550, F-67000 Strasbourg, France}

   \date{Received August 7, 2017; Accepted September 6, 2017}

  \abstract
  {A handful of active galactic nuclei (AGN) have shown strong spectral variations in the optical band
  between epochs that are years apart. The appearance or disappearance of broad emission lines in their spectra 
  completely changes their classification. Since their nucleus orientation cannot change in such short timescales 
  another physical interpretation has to be found. Several scenarios are competing to explain their changing-look 
  nature and, for the first time, we conduct polarized radiative transfer Monte Carlo simulations for all the 
  models. We demonstrate that all interpretations have distinctive features in both total optical flux and 
  continuum polarization such as proposed by Hutsem{\'e}kers and collaborators. Distinguishing between 
  the different scenarios is thus straightforward. We apply our results on the changing-look quasar J1011+5442 
  and confirm the conclusions found by Hutsem{\'e}kers and collaborators: in this specific case, the disappearance 
  of the broad emission lines is due to a change in accretion rate.}


\maketitle

\section{Introduction}
\label{Introduction}
Active galactic nuclei (AGN) are parsec-scale powerful lighthouses that reside in the core of galaxies. 
Their intrinsic luminosity often outshines the light from their host galaxy and such amount of radiation 
can only be produced by accretion onto a supermassive black hole \citep{Pringle1972,Shakura1973}. This 
central engine irradiates its close environment which, in the first half parsec, is mainly constituted 
of electrons and atoms. Through photoionization by continuum photons, emission lines emerge from the 
gas that has a compact and equatorial morphology \citep{Gaskell2009}. Submitted to Keplerian motion, 
this region naturally emits broadened lines. However not all AGN show broad emission lines, which lead 
to the classification of quasars\footnote{Quasar is the historical term coined when the first 
quasi-stellar objects of this type have been discovered. Unlike Seyfert galaxies, quasars are radio-loud, 
cosmological AGN situated in the distant Universe.} according to the presence (type-1) or absence (type-2) 
of those specific spectral features \citep[see, e.g,][]{Rowan-Robinson1977}. We only realized that all 
AGN types might be correlated by their inclination when \citet{Antonucci1985} discovered broad emission 
lines in the polarized flux of a type-2 Seyfert galaxy. It follows that type-1 AGN are seen from the 
polar direction, where there is no equatorial obscuration, while type-2 AGN are seen from the equatorial 
plane. The broad line region (BLR) is thus hidden behind an opaque wall of dust for the later type. The 
classification has several intermediate categories ranging from 1.2 to 1.8, all showing broad emission 
lines, with an increasing [O~{\sc iii}]$\lambda$5007/H$\beta$ flux ratio \citep{Osterbrock1977}. The 
last type, 1.9, includes AGN which only show broad H$\alpha$ lines. 

Once optically classified, an AGN is not expected to flicker between two extreme types, at least 
not in a human timescale. Nevertheless, several ``changing-look'' AGN were discovered in the past decades.
Examples of those AGN are Mrk~1018 \citep{Cohen1986}, NGC~1365 \citep{Risaliti2000}, SDSS~J015957.64+003310.5 
\citep{LaMassa2015} or IC~751 \citep{Ricci2016}. Those AGN are characterized by rapid (month to years 
timescales) dimming/brightening in total flux, and they often show variation in the strength of their
broad emission lines. In several cases the broad emission disappears and an AGN changes from type-1 
to type-1.9 or 2 \citep{Cohen1986}. The disappearance of the broad emission lines are often associated 
with a drop in the continuum flux, suggesting a common physical mechanism \citep{LaMassa2015}. Several
scenarios are proposed to explain the optical-ultraviolet-X-ray variations. First is a possible 
obscuration of the central source by a large amount of dust or gas crossing the observer's line-of-sight. 
This could be due to the motion of dense cloudlets originating from the torus \citep[e.g.,][]{Goodrich1989}
or a variation in the height of the circumnuclear dust region \citep[e.g.,][]{Simpson2005}. This variation 
can occur after an peak in accretion activity and the resulting intense radiation field wipes out the 
outer layers of dust. Conversely, the dusty torus can also recover from this activity and restore its 
amount of dust. A different interpretation was suggested by \citet{LaMassa2015} based on the pioneering 
work of \citet{Elitzur2014}. In this scenario, a change in ionizing flux from the central engine, related 
to a smaller accretion activity, may cause the broad emission features and the BLR to disappear. 
A fainter continuum ultimately reduces the emission line intensity of the BLR since there is not enough 
photons to ionize the gas. The BLR (and later the torus) should disintegrate at small bolometric 
luminosities (L$_{\rm bol} \le$ 5$\times$10$^{39}$ M$^{2/3}_7$, \citealt{Elitzur2009}) as a low 
accretion efficiency would not be able to sustain the required cloud flow rate \citep{Elitzur2006,Elitzur2014}.
Understanding the physics behind the spectral variations of changing-look AGN is thus fundamental to 
probe the life cycle of galaxies, how AGN evacuate or replenish their gaseous/dust material, and 
what is the true morphology of the innermost AGN regions.

To solve these questions, we present radiative transfer simulations of the polarized optical light emerging 
from those different scenarios. We aim to check whether optical polarimetry can unveil the true 
interpretation behind the type-1 to type-1.9/2 changing-look appearance of those peculiar AGN. 
In Sect.~\ref{Modeling} we introduce the Monte Carlo code and the models to be investigated. We explore 
the total flux and polarimetric signal of the different models in Sect.~\ref{Results} and apply our results 
to a specific object in Sect.~\ref{Discussion}. We conclude in Sect.~\ref{Conclusion} about the importance 
of polarimetry to elegantly unveil the correct case-by-case physical interpretation.

\section{Modeling changing-look AGN}
\label{Modeling}
To model the optical continuum polarization expected from a changing-look AGN, we use the 
Monte Carlo radiative transfer code {\sc stokes} \citep{Goosmann2007,Marin2012,Marin2015}.
This numerical tool was used to model, predict, fit and interpret the polarization signatures
of large variety of sources, from exoplanets to AGN \citep[see, e.g.,][]{Marin2014,Marin2017}.
{\sc stokes} works from the near-infrared to the hard X-ray bands and simulates the 
random and successive interactions of light with matter under a large number of possible 
geometries and scales. The code allows to virtually explore all kinds of geometrical 
configurations for an AGN, from the central supermassive black hole to extended
outflows. All the scattering physics is included in the code \citep[see][]{Goosmann2007}
and the user can register the linear and circular polarization at all polar and azimuthal 
viewing angles. The code, in its basic version, is available on-line at: \textcolor{blue}{http://www.stokes-program.info/}.
It is the same version of {\sc stokes} that was used in this publication.

\begin{figure}
   \centering
   \includegraphics[trim = 0mm 70mm 0mm 0mm, clip, width=9cm]{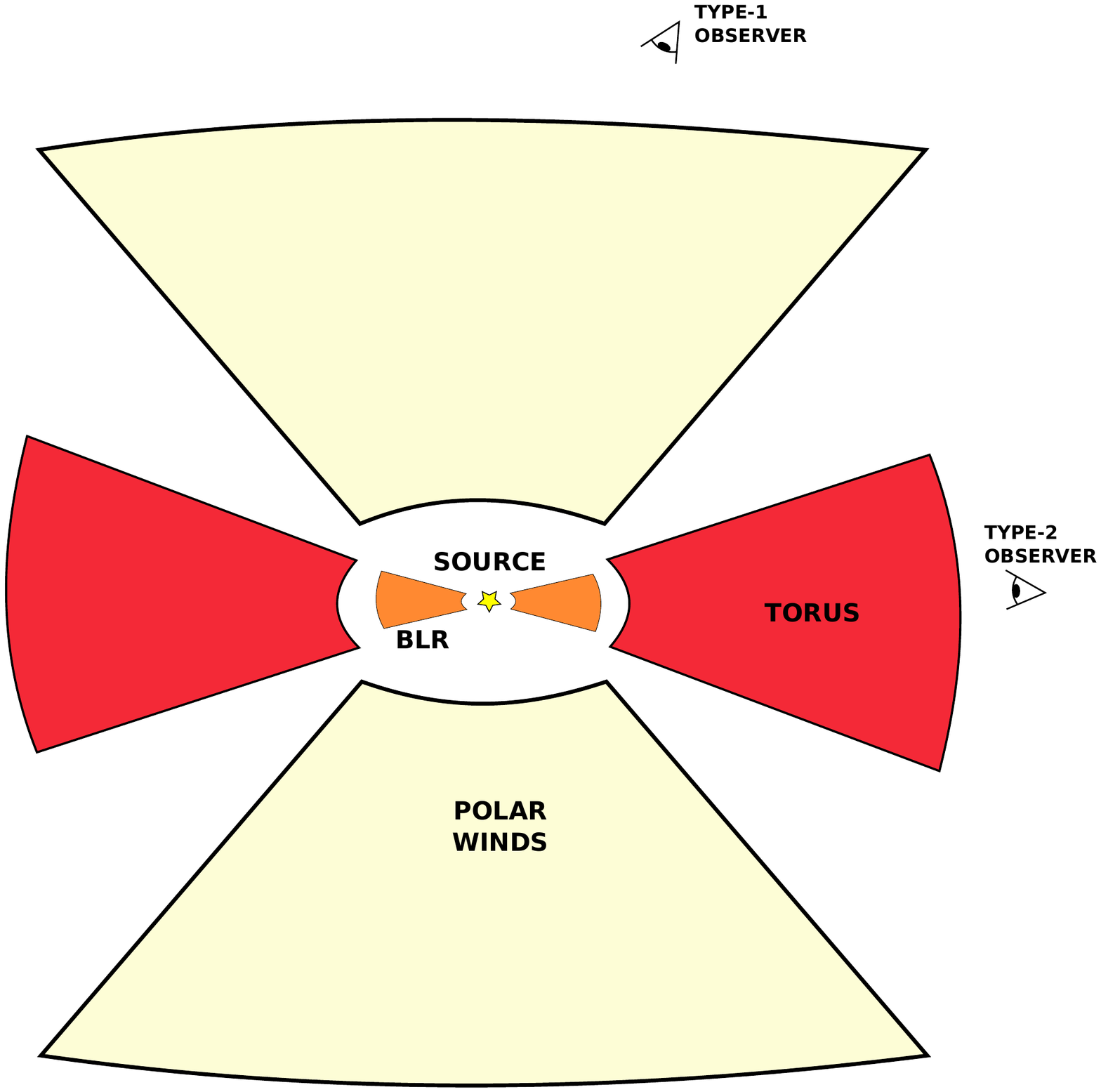}
   \caption{Unscaled AGN model used in this paper. The central optical source 
	    (yellow star) irradiates an electron-filled broad line region 
	    (BLR, in orange) and a coplanar, more distant, dust-filled
	    circumnuclear region (torus, in red). The source also irradiates 
	    torus-collimated, ionized, ejection winds along the polar axis 
	    (primrose yellow). Quantitative details such as sizes, 
	    half-opening angles or composition are given in the text.}
  \label{Fig:Scheme_BLR}
\end{figure}

Our basic AGN model is presented in Fig.~\ref{Fig:Scheme_BLR} (unscaled). The model 
consists of a central point like, isotropic, mono-energetic source that irradiates 
$\lambda$550~nm photons. We selected this wavelength according to the effective wavelength 
midpoint for a standard V filter \citep{Binney1998}. This is also the waveband where a 
large number of historical AGN polarization measures were achieved \citep[see, e.g.,][]{Brindle1990,Smith2002}.
The scattering-induced polarization of AGN being almost-wavelength independent 
in the ultraviolet, optical and near-infrared bands \citep{Miller1983,Code1993,Smith1997}, 
our modeling is thus conservative for broadband observations. Only the amount of 
depolarization from external sources can vary with wavelength, allowing for easier 
polarimetric detections in the ultraviolet waveband where the host contamination is 
less important \citep{Zirbel1998}.

At a distance of 0.01~pc from the continuum source, we set up the inner radius of the 
broad line region according to the theoretical relation between the BLR inner radius 
and the monochromatic flux at $\lambda$550~nm for a 10$^{44}$~erg.s$^{-1}$ AGN luminosity 
\citep{Czerny2015}. The BLR geometry is a flared disk with half-opening angle of 20$^\circ$ 
from the equatorial plane \citep{Marin2012}. It is uniformly filled with electrons and 
the Keplerian motion of the BLR is fixed to 4000~km.s$^{-1}$ \citep{Netzer1990}. The 
radial optical depth in the V-band of the BLR region is set to 3 at the present stage but 
will vary in the photoionizing continuum dimming scenario. The outer radius of the BLR is 
set at the inner radius of the dusty torus, physically constrained by the dust sublimation 
radius. Following the monitoring observations made by \citet{Suganuma2006} in the optical 
and near-infrared wave bands for a set of Seyfert-1 galaxies, we choose a torus inner 
radius of 0.1~pc. The torus outer radius is fixed to 5~pc, following recent interferometric 
observations \citep[see, e.g.,][]{Tristram2007}. The half-opening angle of the circumnuclear 
region is set to 45$^\circ$ and its optical depth is larger than 50 in the V-band. Note 
that the value of the torus half-opening angle will also change in the scenario that 
explains the spectral variations of changing-look quasars by dust obscuration. 
Finally, the torus half-opening angle collimates a pair of conical ejection winds along 
the polar axes. The winds onset at a radial distance of 0.01~pc from the central source 
and stops at 10~pc. It is filled with electrons \citep{Antonucci1985} and its radial 
optical depth is of the order of 0.1 \citep{Marin2012}. This AGN model will be used to 
infer the optical linear continuum polarization of changing-look AGN according to the 
scenarios detailed in Sect.~\ref{Introduction}. For clarity purpose, we focus 
our modeling on type-1 to type-1.9/2 transitions for the remained of this paper.

\section{Results}
\label{Results}

\subsection{AGN polarization and ionizing continuum dimming}
\label{Results:generic}

\begin{figure}
   \centering
   \includegraphics[trim = 10mm 5mm 0mm 0mm, clip, width=10.8cm]{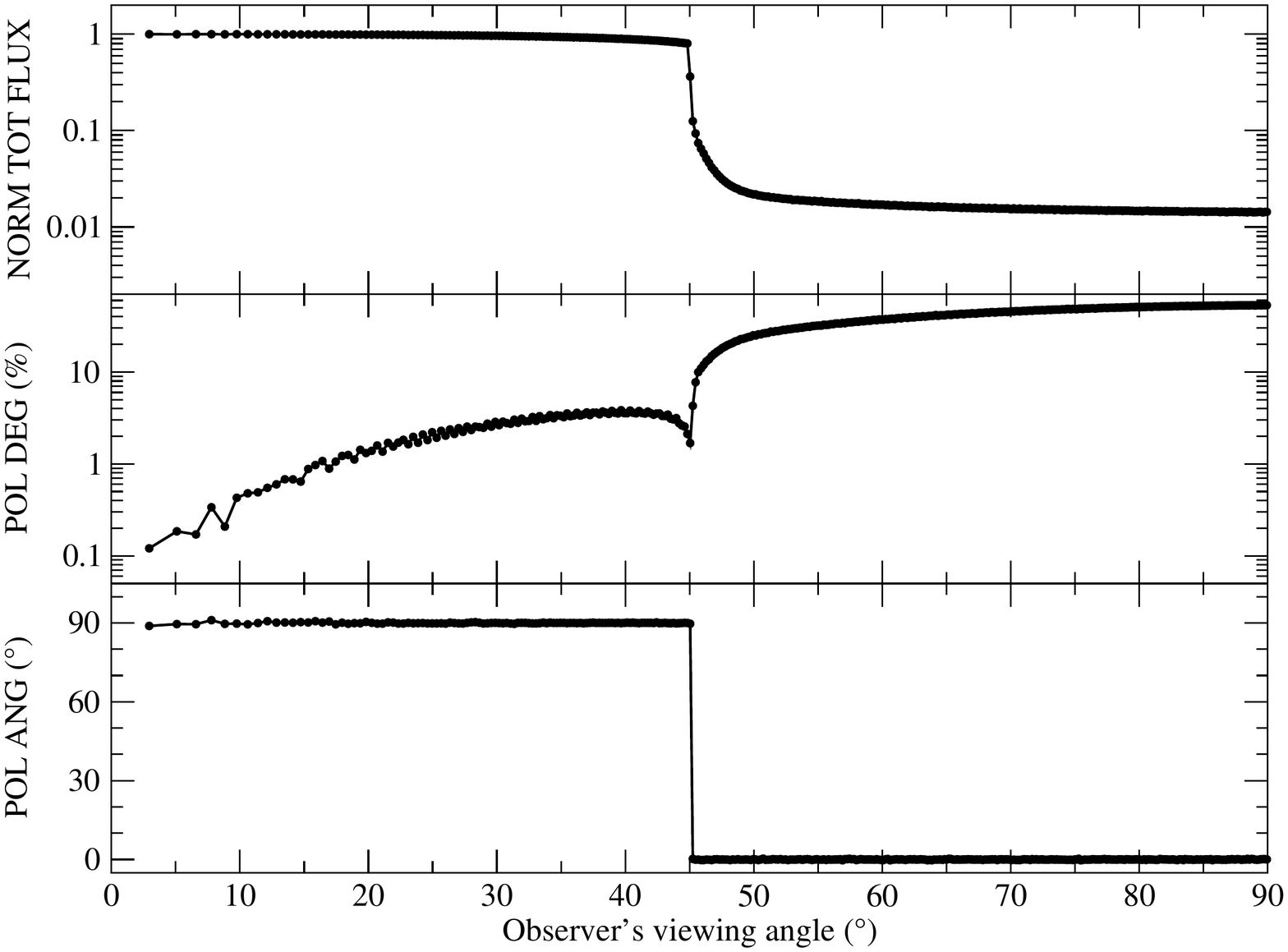}
   \caption{Variation of the optical total flux (normalized, top),
	    optical linear polarization degree (middle) and polarization
	    position angle (bottom) with respect to the inclination 
	    of the observer.}
  \label{Fig:Results_Normal}
\end{figure}

We present the results of our computations for the generic AGN model in Fig~\ref{Fig:Results_Normal}.
The top panel is the total optical flux normalized to the maximum flux observed at an inclination 
of 0$^\circ$. We see that, for type-1 viewing angles (up to the torus horizon, i.e. 45$^\circ$),
the flux is almost constant but suddenly drops at types-2 orientations. This transition 
is also visible in the middle and bottom panels, the continuum linear polarization degree
and polarization position angle, respectively. At type-1 views, the degree of polarization is small
($<$~4\%) and associated with a polarization angle of 90$^\circ$ (parallel to the symmetry axis 
of the system). The polarization degree is null at perfect pole-on inclinations due to the symmetry 
of the system. The polarization degree then increases with inclination until the transition between 
type-1 and type-2 line-of-sights. The polarization position angle rotates down to 0$^\circ$ 
(perpendicular to the symmetry axis of the system) and the polarization degree increases up to 54\%
at equatorial views. In the case of a dimming of the continuum source, the amount of radiation 
scattering of the BLR region will be smaller and the flux will drop. This will result in 
a lower normalized flux at all inclinations and the lack of photoionizing radiation will prevent 
the emission of strong broad emission lines. However it will have no impact on the intrinsic 
polarization degree and angle since they are relative quantities. For this interpretation, 
we expect to see no variations in the (linear continuum) polarization properties of the source.

\subsection{BLR-fading scenario}
\label{Results:BLR}

In the case of a strong dimming (bolometric luminosity $\le$∼10$^{42}$erg.s$^{-1}$), the 
work of \citet{Elitzur2014} suggests that the broad line emission region follows an evolutionary 
sequence that is directly related to the accretion rate of the compact source. If so, the BLR becomes 
less dense as the accretion rate is decreasing, eventually disappearing \citep{Elitzur2006}. 
To investigate this interpretation, we decrease the optical depth of the BLR from 3 to 0.01 
to simulate the progressive disintegration of the region. Following \citet{Elitzur2006}, the timescales 
for the BLR and torus disappearance is short, i.e a few Keplerian orbits. Note that, in our modeling, 
we consider a strong coupling between the BLR and the electron-scattering disk predicted by \citet{Smith2002}.
The equatorial electron region is necessary to reproduce both the observed polarization position angle 
of type-1 AGN and the intrinsic polarization of broadened lines. The electron disk is considered to be 
a continuous flow between the torus and the inner parts of the BLR ($ibid.$). In the work of 
\citet{Smith2002}, the two regions were disconnected since their code only considered single scattering 
but there is no physical reasons for decoupling of the two regions when multiple scattering is enabled. 

\begin{figure}
   \centering
   \includegraphics[trim = 10mm 5mm 0mm 0mm, clip, width=10.8cm]{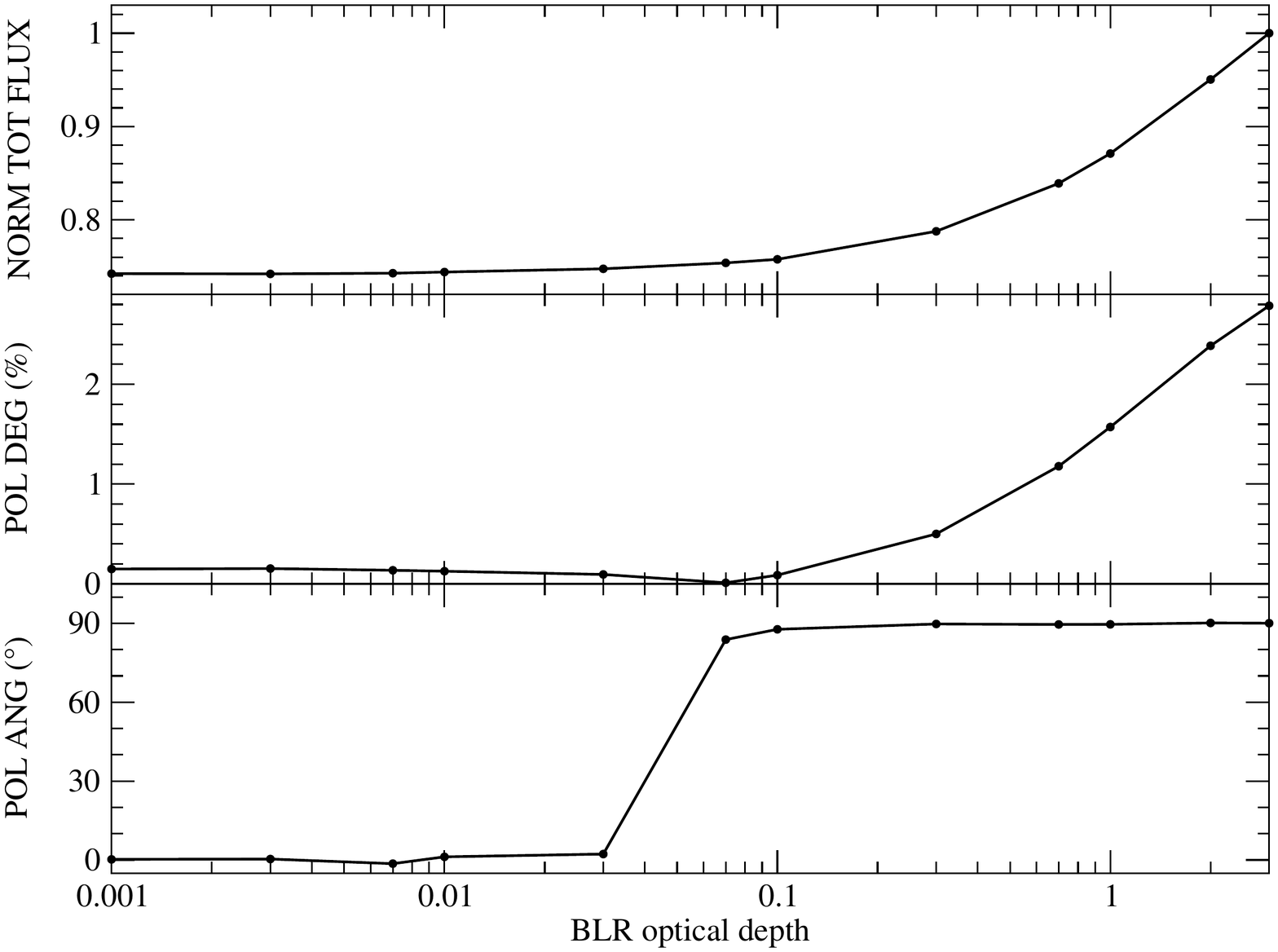}
   \caption{Variation of the optical total flux (normalized, top),
	    optical linear polarization degree (middle) and polarization
	    position angle (bottom) with respect to the gradual fading 
	    of the BLR region. The observer's inclination is set to 30$^\circ$}
  \label{Fig:Results_BLR}
\end{figure}

Fig.~\ref{Fig:Results_BLR} shows the normalized total flux at $\lambda$550~nm, the polarization 
degree and the polarization position angle of the AGN model as a function of the BLR optical depth 
$\tau$. The observer's inclination is fixed at a typical type-1 inclination angle of 30$^\circ$ with 
respect to the symmetry axis of the model. At $\tau$ = 3, the degree of polarization is maximum 
(about 2.79\%) and the polarization position angle is equal to 90$^\circ$. This corresponds to the 
expected polarization angle of type-1 AGN and the relatively high degree of polarization obtained is 
due to the optical thickness of the BLR. The efficiency of the equatorial medium to intercept radiation 
and scatter it towards the observer is maximal. The resulting polarization degree, despite being diluted 
by the unpolarized central source seen by transmission through the wind, is thus superior to 1\%.
For higher densities of the BLR, multiple scattering will start to depolarize radiation. However,
when the BLR region starts to decrease in density due to a lower accretion rate, less electron 
equatorial scattering is happening. The polarization degree thus decreases gradually with $\tau$.
It reaches a minimum around $\tau$ = 0.05 where the polarization position angle rotates from 90$^\circ$
to 0$^\circ$. Equatorial scattering is then inefficient to produce the observed polarization angle 
of Seyfert-1s and the polarization degree remains below 0.15\% for the lower optical depths of 
the BLR. The observed total flux of the AGN follows the same trend as the polarization degree, 
decreasing by 25\% when the BLR is extremely optically thin. We thus see that, if the changing-look 
nature of the few AGN observed is due to the disappearance of the BLR, we expect a strong diminution 
of the total flux. The polarization degree should also decrease and an orthogonal flip of the 
polarization position angle is expected in the optical band. 

\subsection{Varying the amount of dusty obscuration}
\label{Results:Torus}

\begin{figure}
   \centering
   \includegraphics[trim = 0mm 70mm 0mm 0mm, clip, width=9cm]{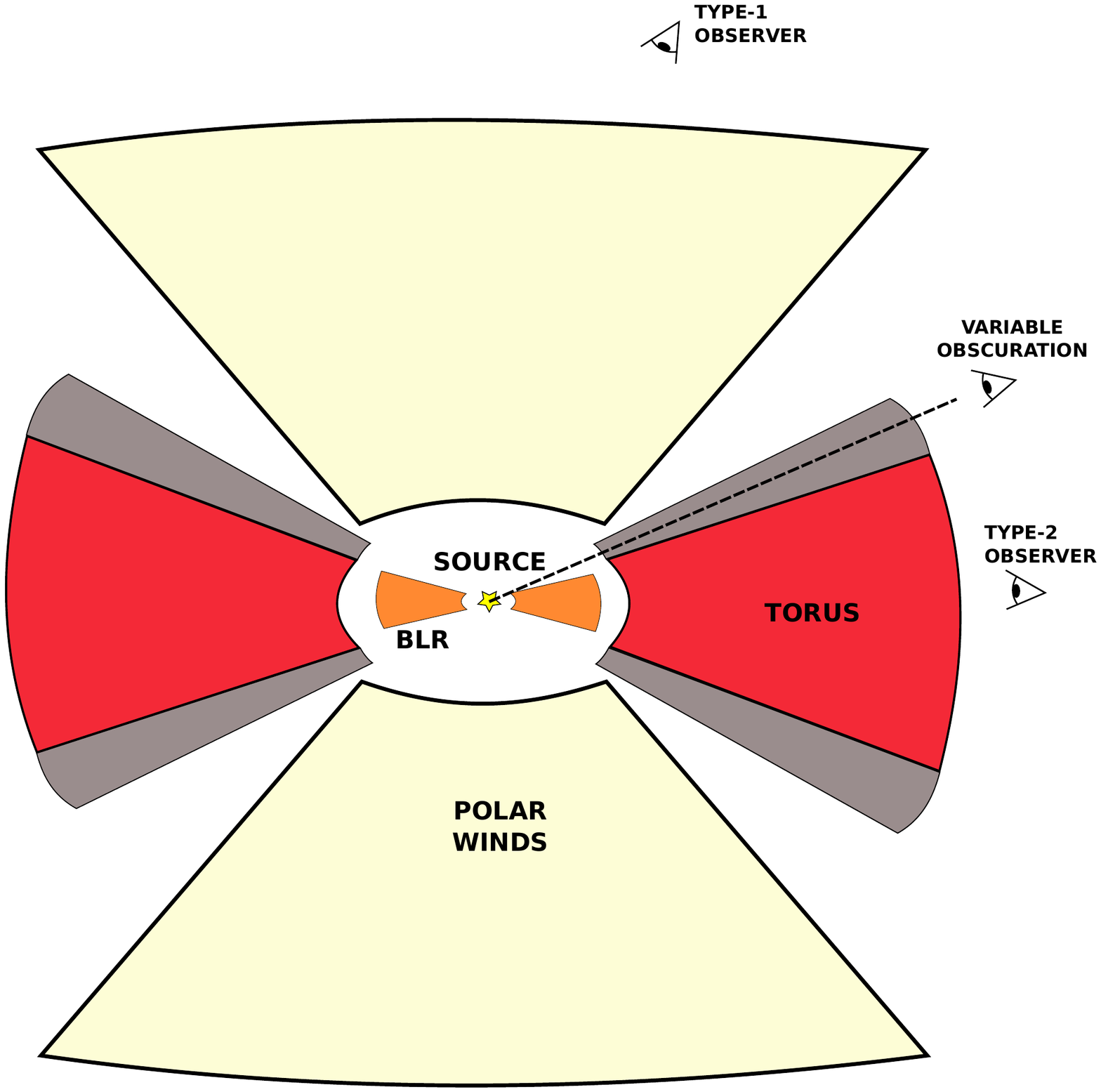}
   \caption{Unscaled AGN model modified according to the variable torus 
	    obscuration interpretation. The outer layers of the 
	    torus (in gray) can be wiped out by an intense radiation 
	    field from the central region or, if the torus is clumpy, 
	    motion of individual clouds located on the observer's 
	    line-of-sight. The spectral type of the AGN thus changes 
	    from type-1.9/2 to type-1.}
  \label{Fig:Scheme_torus}
\end{figure}

\begin{figure}
   \centering
   \includegraphics[trim = 10mm 5mm 0mm 0mm, clip, width=10.8cm]{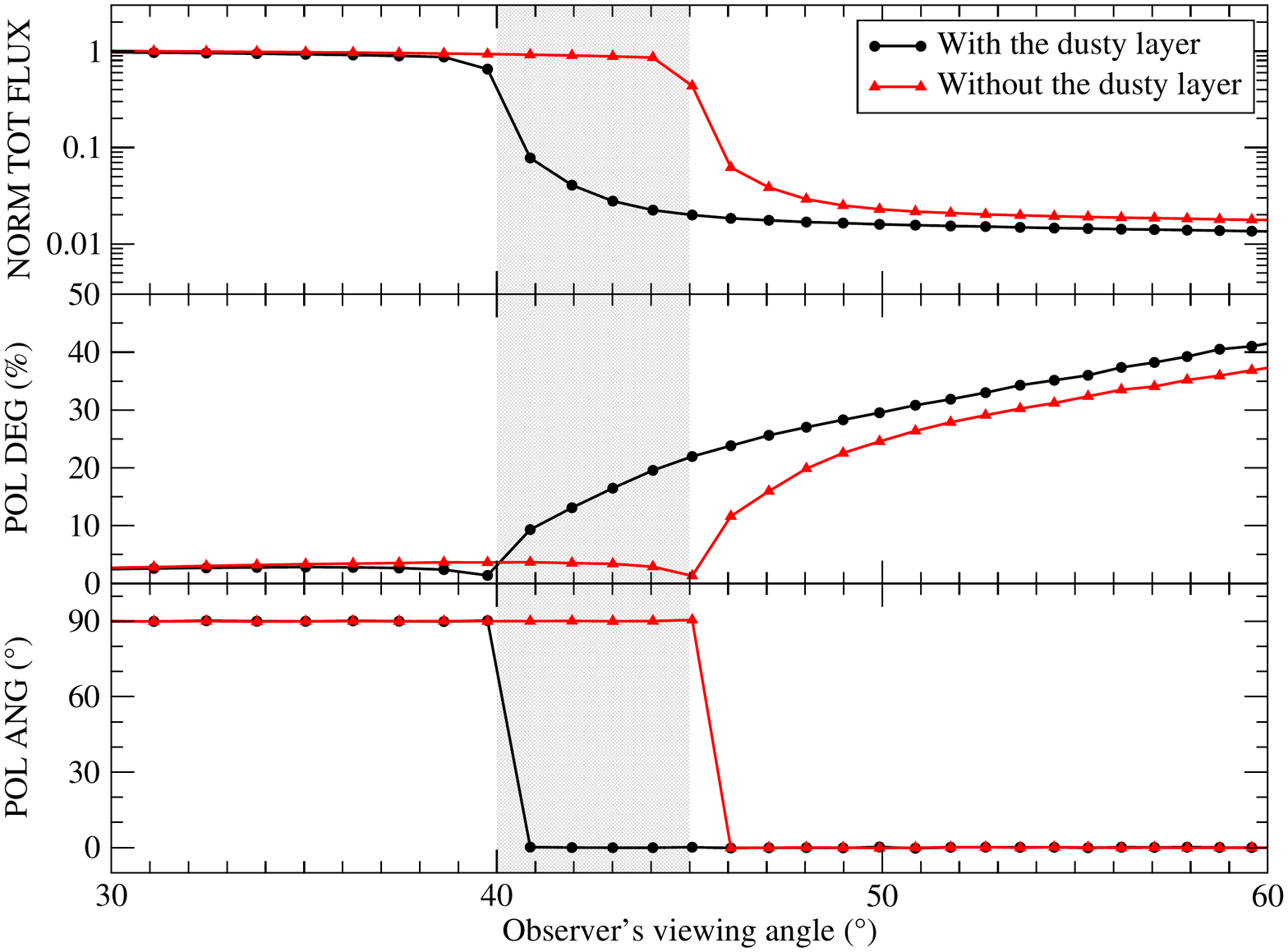}
   \caption{Results of the two scenarios described in 
	    the text: with a dusty layer on top of the torus 
	    (black dots) and when the layer is removed 
	    (red triangles). The top panel shows the normalized 
	    total flux, the middle panel is the polarization 
	    degree, and the bottom panel is the observed polarization 
	    position angle for the two cases. The grey box represents
	    the angular region obscured by the removable dusty layer.}
  \label{Fig:Results_Torus}
\end{figure}

Fig.~\ref{Fig:Results_Torus} presents the results of our modeling for the alternative interpretation
behind the changing-look nature of AGN. In this case, it is believed that the amount of dust 
absorption along the observer's line-of-sight is varying \citep{Matt2003}. To model it, we fix 
the optical depth of the BLR to 3 and add a removable outer layer with an opening angle of 5$^\circ$ 
on the top of the circumnuclear region that is blocking the view of a type-1.9/2 observer. When 
the dusty layer is on the top of the torus, the half-opening angle of the circumnuclear dust material 
changes from 45$^\circ$ to 50$^\circ$ (measured from the equatorial plane). Consequently, the half-opening 
angle of the polar winds is reduced by 5$^\circ$ with respect to the model presented in the previous section. 
It value is then fixed to 40$^\circ$. The layer of dust intercepting the observer's line-of-sight can 
be added or removed at will (see the gray region in Fig.~\ref{Fig:Scheme_torus}). By shaving off the outer 
layers of the torus, we allow a type-1.9/2 observer to have a direct (type-1) view of the central AGN engine 
without changing the nucleus orientation. This is represented by the dashed line in Fig.~\ref{Fig:Scheme_torus}. 

We examine the model from 30 observer's inclinations equally distributed in cosine angle between 
30$^\circ$ and 60$^\circ$. We focus on this range of inclinations as it represents the estimated transition 
angles between type-1 and type-2 AGN categories, slightly depending on the radio-loudness of the quasars 
\citep{Baldi2013,Marin2014b,Sazonov2015,Marin2016,Marin2016b}. Fig.~\ref{Fig:Results_Torus} presents the 
results of the two scenarios: a), when the dusty layer is present on the top of the torus (in black) and b), 
when this layer is removed (in red). The top panel shows the difference in normalized total flux 
between the two scenarios as a function of the observer's inclination. We see that, at the inclinations 
where the observer's viewing angle crosses the removable layer of dust (shaded grey region on 
Figs.~\ref{Fig:Scheme_torus} and \ref{Fig:Results_Torus}), if there is a variable amount of obscuration, the 
flux drops by a factor that is inclination dependent. At maximum, the flux difference is a factor 40 between 
the unobscured (type-1) and obscured (type-1.9/2) model. The transition angle at which the two models change 
from type-1 to type-2 strongly depends on the presence/absence of the dusty layer. At a given inclination, 
one can see the impact of variable obscuration onto the computed flux. The smaller flux observed in the 
type-1.9/2 scenario is due to radiation scattered onto the polar winds, plus a minor contribution of 
backscattering of photons onto the torus funnel opposite to the observer's side. This change in scattering
geometry (from direct light to polar-scattered radiation) has a profound impact on the polarization degree
and angle. For a given inclination within the grey area (where variable obscuration occurs), the 
difference in terms of optical continuum polarization for the two scenarios is large. If the line-of-sight 
toward the nucleus is uncovered, the polarization degree is only a couple of percent, while it rises to 
10 -- 20\% when the torus horizon is obscured with dust. The exact variation of polarization degree between 
the two states of the changing-look AGN depends on the inclination of the observer. Nevertheless, it is 
always a significant change than can be as high as 20\%. This increase in polarization is directly due to 
the different paths radiation has to follow to escape the type-1 and type-1.9/2 models. Additionally, the 
change in spectral type is systematically associated with an orthogonal rotation of the polarization position 
angle.

\section{Discussion}
\label{Discussion}
We saw that the three interpretations (the ionizing continuum dimming, the BLR-fading and the 
variation of dusty obscuration scenarios) have very clear and distinctive features in total flux 
and polarimetry. In the first case, a change in continuum radiation due to a variation in accretion 
rate will not change the observed polarization properties of the AGN. If the accretion rate becomes 
insufficient to sustain the BLR (and torus), the total flux of the AGN should change by a factor 
$\sim$~25\% simply because of the importance of equatorial scattering that redirect photons towards 
the observer. With decreasing accretion rates onto the central supermassive black hole, the polarization 
degree is also decreasing as electron scattering inside the BLR becomes inefficient. The broad line 
emission will appear increasingly weaker and the polarization position angle finally rotates by 90$^\circ$. 
In the last case, the difference in total flux before and after a change of look is much higher. Up to 
98\% of the central flux is absorbed by the circumnuclear dusty medium. Radiation escapes from the AGN 
by scattering inside the polar outflows and thus carry a larger polarization degree due to the Thomson 
laws. The polarization position angle also rotates between the two spectral states. An increase of 
about 10 -- 20\% in polarization degree is expected for a changing-look quasar alternating from a type-1 
to a type-2 classification.

\subsection{The case of J1011+5442}
\label{Discussion:Hutsemekers}
The three interpretations are thus distinctively different, except for their polarization position angle that 
might rotate in the last two cases. By monitoring the flux and polarization state of a sample of changing-look
AGN candidate, it would be possible to easily distinguish the correct physical behind the spectral variations. 
However we do not have archival, large monitoring campaigns of AGN polarization yet. To overcome 
this lack of data, \citet{Hutsemekers2017} used a single polarization measurement as a diagnostic tool of 
the changing-look nature of the quasar SDSS~J101152.98+544206.4 (hereafter J1011+5442). The authors used the 
William Herschel telescope to measure the polarization of J1011+5442, a $z$ = 0.246 quasar with an absolute 
magnitude M$_{\rm i}$ = -22.87 \citep{Shen2011}. Between 2002 and 2015, the blue continua and broad optical 
emission lines of J1011+5442 have been observed to decline \citep{Runnoe2016}, changing the optical classification
of the quasar from type 1 to type 1.9. \citet{Hutsemekers2017} observed J1011+5442 in its faint state 
on February 19, 2017 and, after correcting for the chromatic dependence of the half-wave plate zero-angle 
and instrumental polarization, found a linear polarization degree of 0.15 $\pm$ 0.22\%. Knowing that the
interstellar polarization towards J1011+5442 is expected to be of the order of 0.1\%, \citet{Hutsemekers2017} 
concluded that the polarization of the quasar is compatible with a null intrinsic polarization\footnote{The 
polarization position angle of J1011+5442 could not be estimated with reasonable accuracy \citep{Hutsemekers2017}.}. 
Based on their results and theoretical polarization arguments, the authors suggested that the quasar was 
seen at an inclination close to the pole and that the undetected polarization degree was a proof for the 
lack of photoionizing continuum. The Monte Carlo simulations achieved in this paper clearly prove the correctness 
of their conclusion: if J1011+5442 changed its spectral state because of additional equatorial 
dust obscuration along the observer's line-of-sight, its intrinsic degree of polarization should be of 
the order of 10 -- 20\% and thus easily detectable against interstellar polarization. Thanks to our 
computations, we can go two steps further. First, if the polarization degree of the quasar is of the order 
of 0.15 $\pm$ 0.22\%, its nucleus inclination is strictly inferior to 9$^\circ$ (see Fig.~\ref{Fig:Results_Normal}). 
It is a conservative estimation of the inclination since it does not depend on the exact half-opening angle or 
optical depths of the model components, but rather on the axisymmetry of the unified model itself \citep{Marin2012}.
Second, looking at Fig.~3 from \citet{Runnoe2016}, the decrease in flux at $\lambda$550~nm between 
the two epoch is approximatively 35\%. This drop in flux is similar to what we found in Sect.~\ref{Results}
when the BLR is fading away (25\%). The difference can be easily compensated by an intrinsic dimming of the
continuum source and/or a modification of the geometrical configuration of the broad emission line region 
(the half-opening angle and optical depth of the BLR being the two critical parameters here). However, this 
drop in flux is clearly not compatible with the 98\% diminution expected in the scenario where dust clouds 
from the torus genuinely block radiation from the central engine.

\subsection{Polarization reverberation mapping}
\label{Discussion:Reverberation}
As already mentioned in \citet{Hutsemekers2017}, scattering in AGN is considered to take place up 
to a few parsecs. Additionally, the amount of reprocessing events is inclination-dependent, such as shown in 
Fig.~\ref{Fig:Results_Normal} and explained in Sect.~\ref{Results:generic}. Hence, if the central continuum 
source suddenly suffers a strong dimming, polarization from the polar outflows which extend over 10~pc 
\citep{Capetti1995} is expected to last up to $\sim$~32 years after the continuum change. A polarimetric 
echo of the past core activity will remain visible in the extended structures of the AGN, such as expected 
in the case of the Galactic center in the X-ray band \citep{Churazov2002,Marin2014c,Marin2015b}. The 
observed polarization properties of an AGN, if integrated over the whole structure, will be impacted.
Since the timescale of changes of look is of the order of a few years \citep{MacLeod2016}, this effect 
might have an impact on the proposed polarization diagnostics unless high angular resolution polarization 
maps are available for nearby objects (such as for the case of NGC~1068 which was observed using the 
Spectro-Polarimetric High-contrast Exoplanet REsearch instrument -- SPHERE -- on the Very Large Telescope, 
see \citealt{Gratadour2015}).

This is where the polarization reverberation mapping technique is the most valuable. \citet{Gaskell2012}
demonstrated how the polarization variability in NGC~4151 can be used to probe the size and structure 
of scattering regions. Between 1997 and 2003, the type-1 radio-quiet AGN NGC~4151 has shown variations 
of an order of magnitude in its optical polarized flux while its polarization position angle remained 
constant. Since the sizes of the different scattering regions in AGN span over several orders of magnitude, 
the time delay we measure is thus directly related to the geometry of the system. Scattering inside the BLR 
produces a polarization angle parallel to the radio axis of the system and its temporal delay is shorter 
than photons that have scattered onto the polar outflows, where the scattering-induced polarization angle 
is perpendicular. It follows that polarization reverberation mapping can precisely locate where scattering 
happens. In addition, if temporal changes in the wavelength dependence of polarization across broad emission 
lines are detected, this could imply a change in the BLR and scatterer geometry \citep{Shoji2005}. This effect 
would be enhanced for increasing nucleus inclinations and the two scenarios discussed in this paper could be, 
in principle, distinguished thanks to polarization reverberation mapping (Rojas Lobos et al. submitted; 
Marin et al. in prep.).

\section{Conclusions}
\label{Conclusion}
In this work, we have investigated the different interpretations behind the changing-look nature 
observed for a few AGN. The first model predicts that the disappearance of the broad emission line and 
the decrease in flux are due to the vanishing of the broad emission line region, linked with a decrease
of the black hole accretion rate. If the accretion rate becomes too low, the BLR itself can progressively 
disappear since mass accretion can no longer sustain the required column densities. The last model 
explains the same spectral changes with a variation in the obscuration of the observer's viewing angle that 
is grazing the torus horizon. Either the outer layers of the torus are puffed-up or clouds intercept the 
line-of-sight. For the first time, both models were investigated using radiative transfer Monte Carlo 
calculations and their optical flux and polarization signals were found to be distinctively different. 
If the spectral variations are due to a lack of photionizing radiation, the flux should drop accordingly 
but no variation of the polarization properties are expected, since they are relative quantities 
and they do not depended on the amount of photons scattering inside the BLR. In the case of a progressive
disappearance of the BLR, a $\sim$~25\% decrease in total flux and a polarization degree $<$~0.1\% are 
expected. Finally, if variable obscuration is the correct scenario, the total flux should be reduced by 
about a factor 40 and the polarization degree should be, at least, 10\% higher. In the last two cases the 
polarization position angle should rotate by 90$^\circ$.

We applied our results to the J1011+5442 quasar investigated by \citet{Hutsemekers2017} and found 
strong evidences for the correctness of their interpretation. We extended the polarimetric investigation
of the authors by estimating the nucleus inclination of the quasar ($<$~9$^\circ$). Our computations 
show that the total flux variation of the quasar over a decade is also consistent with their primary 
conclusion: the broad emission line region of J1011+5442 switched off between 2002 and 2015. The 
same conclusion was postulated by \citet{LaMassa2015} and \citet{MacLeod2016} based on specific 
and systematic photometric searches for changing-look quasars. We, for the first time, provide detailed 
computations of the expected flux attenuation and optical polarization expected from all the 
scenarios. 

Our paper clearly shows that distinguishing between the various physical interpretations is easy. New 
optical polarimetric observations of AGN showing a changing-look behavior will immediately tell if the 
change is due to an intrinsic dimming of the ionizing continuum source, a BLR disappearance or a variation 
in line-of-sight obscuration. Prior polarimetric measurements are not vital since the actual intrinsic 
polarization degree should be significant enough to tell apart the two last models (null or $<$~0.1\% 
polarization in the first case, high, $\ge$~10\% polarization in the other case). Archival polarimetric 
information about the polarization angle and degree would be beneficial to detect if the spectral change 
is simply due to a lack of photoionizing radiation (no variations in polarization between the two epochs) 
or a progressive disappearance of the BLR (decrease of polarization degree and rotation of the polarization 
position angle). Coupling the polarimetric measurements with past photometric data would greatly facilitate
the interpretation, since the change in flux level is also model-dependent. We thus advocate for systematic 
polarimetric observations of changing-look AGN in order to fully understand their true nature.

\acknowledgements 
The author would like to acknowledge the anonymous referee for her/his useful comments that improved 
the quality of the paper.

\bibliographystyle{aa}
\bibliography{biblio}

\end{document}